\documentclass[12pt, a4paper]{amsproc}

\usepackage{amsmath, amssymb, amsthm} 

\newcommand{\cprime}{\/{\mathsurround=0pt$'$}}

\newtheorem{theorem}{Theorem}
\newtheorem{proposition}{Proposition}
\newtheorem{lemma}{Lemma}

\theoremstyle{definition}

\newtheorem{remark}{Remark}

\begin{document}

\author{Sergei Igonin}
\address{Sergei Igonin \\ 
Independent University of Moscow, Russia  
and University of Twente, the Netherlands}
\email{igonin@mccme.ru}

\author{Ruud Martini}
\address{Ruud Martini \\ University of Twente \\ 
Faculty of Mathematical Sciences \\
P.O. Box 217 \\ 7500 AE Enschede\\ the Netherlands}
\email{martini@math.utwente.nl}

\title[Prolongation of the Krichever-Novikov equation]
{Prolongation structure of the Krichever-Novikov equation}


\begin{abstract}
We completely describe  
Wahlquist-Estabrook 
prolongation structures (coverings)
dependent on $u,\,u_x,\,u_{xx},\,u_{xxx}$
for the Krichever-Novikov equation
$u_t=u_{xxx}-3u_{xx}^2/(2u_{x})+p(u)/u_{x}+au_{x}$
in the case when  
the polynomial $p(u)=4u^3-g_2u-g_3$ has distinct roots. 
We prove that there is a universal prolongation algebra isomorphic 
to the direct sum of a commutative $2$-dimensional algebra and a
certain subalgebra of the tensor product of 
$\mathfrak{sl}_2(\mathbb{C})$ with the algebra of regular functions 
on an affine elliptic curve.
This is achieved by identifying this prolongation algebra 
with the one for the anisotropic Landau-Lifshitz equation. 
Using these results, 
we find for the Krichever-Novikov equation  
a new zero-curvature representation, which is 
polynomial in the spectral parameter 
in contrast to the known elliptic ones.


\end{abstract}

\maketitle

\section{Introduction}

The Krichever-Novikov (KN) equation
\begin{equation}
  \label{kn}
 u_t=u_3-\frac32\frac{u_2^2}{u_1}+
 \frac{4u^3-g_2u-g_3}{u_1}+au_1,\quad
 u_k=\frac{\partial^k u}{\partial x^k},\quad 
  g_2,\,g_3,\,a \in\mathbb{C},  
\end{equation}
appeared for the first time in \cite{krich} in connection with a study
of finite-gap solutions of the KP equation. If the roots 
$e_1,\,e_2,\,e_3$ of the polynomial $4u^3-g_2u-g_3$ are 
distinct then equation \eqref{kn} is called \emph{nonsingular}. 
According to \cite{sok1,sok2}, in this case 
no differential substitution 
\begin{equation}
\label{miura}
\tilde u=g(u,u_1,u_2,\dots) 
\end{equation}
exists connecting \eqref{kn} 
with other equations of the form 
\begin{equation}
  \label{kdv_type}
  u_t=u_3+f(u,u_1,u_2).
\end{equation}
Moreover, nonsingular equations \eqref{kn} exhaust  
(up to invertible transformations $u=\varphi(\tilde u)$) 
all the \emph{integrable} 
(possessing an infinite series of conservation laws) 
equations \eqref{kdv_type} that are not reducible 
by a finite number of substitutions \eqref{miura} to the KdV equation
$u_t=u_3+u_1u$ or the linear equation $u_t=u_3+au_1$.

These distinctive features make equation \eqref{kn} worth to study in detail. 
In this paper we apply the Wahlquist-Estabrook prolongation method
to it. Some particular zero-curvature representations 
\cite{loop,krich,novikov} as well as a 
B\"acklund transformation \cite{adler} 
for \eqref{kn} are known, but 
a complete description of prolongation structures has not been given,
and we perform this below. 

It turns out that with respect to 
prolongation structures equation \eqref{kn} 
continues to demonstrate remarkable properties. First of all, 
in order to obtain nontrivial results one has to consider
prolongation structures of order $3$ 
(i.e., dependent on $u_k,\,k\le 3$) 
in contrast to the normal assumption that their order is 
lower than the equation's order. 
Because of this, there is additional gauge freedom,
which impedes the computation. Fortunately, 
we find a canonical form for the considered prolongation structures, 
which fixes partially the gauge
and makes it possible to obtain a universal prolongation algebra in
terms of generators and relations. In Section \ref{cov-kn} 
we show that this is in fact the case for any equation of the form  
$u_t=u_3-3u_2^2/(2u_1)+f(u)/u_1+au_1$.

Following \cite{nonl,rb}, in order to clarify the computation and the nature
of gauge transformations we interpret differential equations as submanifolds
in infinite jet spaces and prolongation structures
as special morphisms called \emph{coverings} of such manifolds.
This method is recalled in Section \ref{psc}.

In \cite{loop, krich} it is noticed that Sklyanin's zero-curvature
representation for the anisotropic Landau-Lifshitz (LL) equation leads 
by means of a special transformation of the dependent variables 
to a zero-curvature representation for the nonsingular equation \eqref{kn}. 
Note that this is not a B\"acklund transformation and does not 
establish any correspondence between solutions of the two equations.
 
In Section \ref{pakn} we make use of this transformation to
choose special generators in the prolongation algebra $\mathfrak{g}$ of 
\eqref{kn} in the nonsingular case such that 
the resulting relations turn into 
the ones for the LL prolongation algebra. 
In \cite{ll} the latter algebra was explicitly described, 
and we recall this description in Section \ref{pall}.

Finally, in Section \ref{pakn} 
we prove that $\mathfrak{g}$ is isomorphic to 
the direct sum of a commutative $2$-dimensional algebra and 
a certain subalgebra of the tensor product of $\mathfrak{sl}_2(\mathbb{C})$ 
with the ring $\mathbb{C}[v_1,v_2,v_3]/\mathcal{I}$, where the ideal $\mathcal{I}$
is generated by the polynomials
\begin{equation*}
v_i^2-v_j^2+\frac83(e_j-e_i),\quad i,\,j=1,2,3, 
\end{equation*}
defining a nonsingular elliptic curve in $\mathbb{C}^3$.

In particular, we establish one-to-one correspondence 
between zero-cur\-va\-tu\-re 
representations (ZCR) for the anisotropic LL equation
and the nonsingular KN equation.
Using this, in Section \ref{pzcr} 
we derive a new $\mathfrak{sl}_4(\mathbb{C})$-valued ZCR 
for the nonsingular KN equation from
the found in \cite{polyn} ZCR for the LL equation.
Remarkably, 
this ZCR is polynomial in the spectral parameter 
in contrast to the known for \eqref{kn} ZCR with elliptic 
parameters \cite{loop,krich,novikov}. 

Generally,
we think that, side by side with the symmetry algebra 
and the space of conservation laws, 
the prolongation algebra is an important invariant of 
a given system of differential equations. 

The obtained algebra $\mathfrak{g}$
differs considerably from other known prolongation algebras
for equations of the form \eqref{kdv_type}. 
Indeed, for the KdV equation and the potential KdV equation 
$u_t=u_3+u_1^2$ the Lie algebras governing the  
prolongation structures of order $2$ 
were described explicitly in
\cite{kdv_pa} and \cite{pkdv} respectively. 
In both cases the algebra turned out to be the direct sum of the
polynomial loop algebra 
$\mathfrak{sl}_2\otimes_{\mathbb{C}}\mathbb{C}[\lambda]$ and a
finite-dimensional nilpotent algebra.

\section{Prolongation structures as coverings}
\label{psc}

We use the following modification due to Krasilshchik and Vinogradov
\cite{nonl,rb} of the original Wahlquist-Estabrook method. 
Let ${\mathcal{E}}\subset J^\infty(\pi)$ 
be the (infinite-dimensional) submanifold determined 
by a system of differential equations 
and its differential consequences in the infinite jet space 
$J^\infty(\pi)$ of some smooth bundle $\pi\colon E\to U$,
where $U$ is an open subset of $\mathbb{R}^n$.
Let $x_1,\dots,x_n$ be coordinates in $U$, 
which play the role of independent variables in the equations.
The total derivative operators $D_{x_i}$ are treated as 
commuting vector fields on ${\mathcal{E}}$.

A \emph{covering over} ${\mathcal{E}}$ is given by a smooth bundle 
\begin{equation}
\label{psi}
\psi\colon\tilde{\mathcal{E}}\to{\mathcal{E}}	
\end{equation}
and an $n$-tuple of vector fields $\tilde D_{x_i},\,i=1,\dots,n$, 
on the manifold $\tilde{\mathcal{E}}$ such that 
\begin{gather}
  \label{proj}
  \psi_*(\tilde D_{x_i})=D_{x_i},\\
  \label{comm1}
  [\tilde D_{x_i},\,\tilde D_{x_j}]=0,\quad \forall\,i,j=1,\dots,n.
\end{gather}
See \cite{nonl} for a motivation of this definition 
and its coordinate-free formulation. 

A diffeomorphism $\varphi\colon \tilde{\mathcal{E}}\to\tilde{\mathcal{E}}$ such
that $\psi\circ\varphi=\psi$ is called a \emph{gauge transformation},
and the covering given by $\{\varphi_*(\tilde D_{x_i})\}$
is said to be \emph{gauge equivalent} to the covering $\{\tilde D_{x_i}\}$.

In this paper we consider equations in two independent variables
$x$ and $t$, i.e., $n=2$. 
Then Wahlquist-Estabrook prolongation structures \cite{prol} 
correspond to the case when \eqref{psi} is a trivial bundle 
$$
\psi_{\mathrm{tr}}\colon{\mathcal{E}}\times W\to{\mathcal{E}},
\quad \dim W=m<\infty.
$$
Local coordinates $w^1,\dots,w^m$ in $W$ correspond 
to \emph{pseudopotentials} in the Wahlquist-Estabrook approach \cite{prol}.  
From \eqref{proj} we have 
\begin{equation}
\label{tilde}
	\tilde D_x=D_x+A,\quad \tilde D_t=D_t+B,
\end{equation}
where   
\begin{equation}
\label{AB}
A=\sum_jA^j{\partial}_{w^j},\quad B=\sum_jB^j{\partial}_{w^j}
\end{equation}
are $\psi_\mathrm{tr}$-vertical vector fields. 
Condition \eqref{comm1} is written as
\begin{equation}
  \label{cov}
  D_x B-D_t A+[A,B]=0.
\end{equation}
A covering gauge equivalent to the one given by 
$\tilde D_x=D_x,\,\tilde D_t=D_t$ is called \emph{trivial}.

We call a $\psi_\mathrm{tr}$-vertical vector field $A$ on ${\mathcal{E}}\times W$ \emph{linear}
(with respect to the given system of coordinates in $W$) 
if $A=\sum_{ij} a_{ij}w^j{\partial}_{w^i}$ for some functions $a_{ij}\in C^\infty({\mathcal{E}})$. 
Denote by $A_M$ the $m\times m$ matrix-function on ${\mathcal{E}}$ 
with the entries $[A_M]_{ij}=a_{ij}$.
For two linear vector fields $A,\,B$ the commutator $[A,B]$ is also linear, 
and one has
\begin{equation}
\label{ABM}
[A,B]_M=[B_M,A_M].
\end{equation}

If the linear vector fields $A,\,B$ meet \eqref{cov} then the matrices
$A_M,\,B_M$ satisfy
\begin{equation}
\label{MN}
[D_x-A_M,\,D_t-B_M]=D_tA_M-D_xB_M+[A_M,B_M]=0.	
\end{equation}
and form a \emph{zero-curvature representation} (ZCR) of ${\mathcal{E}}$. 
The functions $A_M,\,B_M$ may in fact take values 
in an arbitrary Lie algebra $\mathfrak{g}$, and 
then the ZCR is said to be \emph{$\mathfrak{g}$-valued}.

\section{Coverings of KN type equations}
\label{cov-kn}

In this section we solve \eqref{cov} for equations of the form 
\begin{equation}
  \label{gkn}
  u_t=u_3-\frac32\frac{u_2^2}{u_1}+\frac{p(u)}{u_1}+au_1,\quad 
  u_k=\partial^ku/\partial x^k,
\end{equation}
where $u$ is a complex-valued function of two real variables $x,\,t$
and $p(u)$ is an arbitrary analytic function of $u$.
In this case 
$\pi\colon\mathbb{C}\times\mathbb{R}^2\to\mathbb{R}^2,\ (u,x,t)\mapsto (x,t)$.

\begin{remark}
Here the bundle $\pi$ and its jet bundles are complex, 
while \cite{nonl,rb} deal with real bundles.
However, it is easily seen that the theory 
of coverings is the same for complex bundles. 
In our case this follows from
the concrete formulas presented below. 
\end{remark}

The manifold ${\mathcal{E}}$ has the natural coordinates 
$x,\,t,\,u_k,\ k\ge 0$, where $x,\,t$ are real and $u_k$ are complex.
The total derivative operators are written in these coordinates as follows 
\begin{align}
  \label{dx}
  D_x&=\partial_x+\sum_{j\ge 0} u_{j+1}\partial_{u_j},\\
  \label{dt}
  D_t&=\partial_t+\sum_{j\ge 0} D_x^j(F)\partial_{u_j},
\end{align}  
where $F$ is the right-hand side of \eqref{gkn}.

Below $w^1,\dots,w^m$ are also complex, 
and all functions and 
vector fields are complex-valued and analytic with respect to
their complex arguments.

Studying coverings over an evolution equation 
$u_t=f(u,u_1,\dots,u_p)$,
one normally assumes to simplify the problem 
that $A,\,B$ in~\eqref{cov} 
do not depend on the variables $x,\,t$ 
and the derivatives $u_k,\,k\ge p$. 
However, in order to obtain nontrivial coverings 
for equation \eqref{gkn} we have to allow $A,\,B$
to depend at least on $u_k,\,k\le 3$ 
(and, of course, on $w^1,\dots,w^m$), 
see Remark \ref{u3} below. 

Then a straightforward computation shows 
that \eqref{cov} requires  
$A=A(w,u,u_1,u_2,u_3)$ to be of the form
\begin{equation}
\label{A2}
A=\frac{1}{u_1}A_1(w,u)+A_0(w,u)+u_1A_2(w,u)  
\end{equation}
Here and in what follows the symbol $w$ stands 
for the whole collection $w^1,\dots,w^m$.
We want to get rid of the term $u_1A_2(w,u)$ 
by switching to a gauge equivalent covering.

To this end, let $A_2(w,u)=\sum_ja^j(w^1,\dots,w^m,u){\partial}_{w^j}$ 
and fix $u'\in\mathbb{C}$.
Consider a local analytic solution 
of the system of ordinary differential equations
\begin{equation}
\label{ode}
\frac{d}{du}f^j(w,u)=a^j(f^1,\dots,f^m,u),\quad j=1,\dots,m,	
\end{equation}
dependent on the parameters $w$ 
with the initial condition $f^j(w,u')=w^j$.

Then the formulas 
\begin{equation}
\label{kill}
  u_k\mapsto u_k,\quad w^j\mapsto f^j(w,u),\quad k\ge 0,\ j=1,\dots,m,
\end{equation}
define a local gauge transformation
$\varphi\colon{\mathcal{E}}\times W\to {\mathcal{E}}\times W$ 
such that $\varphi_*(D_x+A)=D_x+A'$, where the vector field 
$A'$ is of the form \eqref{A2} without the linear in $u_1$ term.

\begin{remark}
This is easily seen from the following interpretation 
of coverings \cite{nonl,rb}. 
The manifold ${\mathcal{E}}\times W$ is itself isomorphic to
the submanifold in an infinite jet space
determined by the system consisting 
of equation \eqref{gkn} and 
the following additional equations 
\begin{equation}
\begin{aligned}
\label{naiv}
	\frac{\partial w^j}{\partial x}&=A^j(w,u,u_1,u_2,u_3),\\
	\frac{\partial w^j}{\partial t}&=B^j(w,u,u_1,u_2,u_3),
\end{aligned}	
\quad \quad j=1,\dots,m,
\end{equation}
where $A^j,\,B^j$ are the components of $A,\,B$
in \eqref{AB}. The vector fields $D_x+A,\,D_t+B$
are the restrictions of the total derivative operators
to ${\mathcal{E}}\times W$.
Gauge transformations correspond to invertible changes of variables
\begin{equation*}
  w^j\mapsto g^j(x,t,w,u,u_1,\dots),\quad j=1,\dots,m,
\end{equation*}
in \eqref{naiv}. Clearly, due to equation \eqref{ode} 
after substitution \eqref{kill} 
the linear in $u_1$ terms contract in \eqref{naiv}. 
\end{remark}

Since we are interested in local 
classification of coverings up to gauge
equivalence, we can from the beginning assume that
\begin{equation}
  \label{Au_1}
  A=\frac{1}{u_1}A_1(w,u)+A_0(w,u).
\end{equation}
\begin{remark}
This rather unusual step in solving \eqref{cov} is due to the
assumption that $A,\,B$ may depend on the derivatives $u_k,\,k\le 3$.  
\end{remark}

Further computation shows that $A_0(w,u)$ does not actually depend 
on $u$. Denote for brevity $A_1=A_1(w,u)$ and $A_0=A_0(w)$. 
Finally, we obtain
\begin{multline}
  \label{b}
  B=-\frac{u_3}{u_1^2}A_1+ 
  \frac{u_2^2}{2u_1^3}A_1+
  \frac{2u_2}{u_1}\frac{\partial A_1}{\partial u}+\frac{u_2}{u_1^2}[A_0,A_1]\\
  -\frac{p(u)}{3u_1^3}A_1+\frac{2}{u_1}[A_1,\frac{\partial A_1}{\partial u}]
  -2u_1\frac{\partial^2 A_1}{\partial u^2}+aA+B_0(w),
\end{multline}
where
\begin{gather}
  \label{2}
  \frac{\partial^3 A_1}{\partial u^3}=0,\\
  \label{0}
  [A_0,B_0]=[A_1,A_0]=[A_1,B_0]=0,\\
  \label{-2}
  2p\frac{\partial A_1}{\partial u}-\frac{\partial p}{\partial u}A_1-
  3[A_1,[A_1,\frac{\partial A_1}{\partial u}]=0.
\end{gather}
\begin{remark}
\label{u3}
From \eqref{b} we see that if $B$ does not depend on 
$u_3$ (i.e., $A_1=0$) then 
$A,\,B$ do not depend on $u_k$ at all and, therefore, 
the covering is trivial. Hence our
assumption that $A,\,B$ may depend on 
$u_k,\,k\le 3$, is essential.
\end{remark}

From \eqref{2} we see that $A_1$ is a polynomial in $u$ 
of degree not greater than $2$, i.e.,
\begin{equation}
  \label{A1}
  A_1=A_{10}+uA_{11}+u^2A_{12}
\end{equation}
for some vector fields $A_{1j}=A_{1j}(w)$.

Thus any covering of the considered type is uniquely determined by 
$5$ independent of $u_k,\,k\ge 0$, vector fields on $W$  
\begin{equation}
  \label{gener}
  A_0,\ B_0,\ A_{1j},\ j=0,1,2,
\end{equation}
subject to restrictions \eqref{0}, \eqref{-2}.
For each concrete function $p(u)$ equations \eqref{0} and 
\eqref{-2} give some relations between vector fields \eqref{gener}. 
As usual, the quotient 
of the free Lie algebra generated by letters \eqref{gener} 
over these relations is called the \emph{prolongation algebra} of
equation \eqref{gkn}. 
From \eqref{0} we see that $A_0,\,B_0$ lie
in the center of the prolongation algebra. 

In Section \ref{pakn} 
we solve these relations in the case when $p(u)$
is a polynomial of degree $3$ with distinct roots. 
To achieve this, the description of the prolongation algebra 
for the Landau-Lifshitz equation is needed, 
which we recall in the next section.

\section{Prolongation structure of the Landau-Lifshitz equation}
\label{pall}

The Landau-Lifshitz (LL) equation reads \cite{loop,ll}
\begin{equation}
  \label{l-l}
  \mathbf{S}_t=\mathbf{S}\times \mathbf{S}_{xx}+\mathbf{S}\times J\mathbf{S},\quad S_1^2+S_2^2+S_3^2=1,
\end{equation}
where $\mathbf{S}=(S_1,S_2,S_3)$ is a complex-valued vector-function of $x,t$ 
and $J=\mathrm{diag}(j_1,j_2,j_3),\,j_k\in\mathbb{C},$ is a diagonal matrix.

For \eqref{l-l} equation \eqref{cov} under the normal 
assumption that $A,\,B$ do not depend on $x,\,t$ and 
derivatives of $\mathbf{S}$ of order $>1$ 
was solved in \cite{ll} as follows 
\begin{align}
 \label{All}
 A&=\mathbf{P}\cdot \mathbf{S}+P_4,\\
 \label{Bll}
 B&=(\mathbf{P}\times \mathbf{S})\cdot \mathbf{S}_x+(\mathbf{P}\times\mathbf{P})\cdot \mathbf{S}+P_5,
\end{align}
where $\mathbf{P}=(P_1,P_2,P_3)$ and
$\mathbf{P}\times\mathbf{P}=([P_2,P_3],[P_3,P_1],[P_1,P_2])$.
Here the vector fields $P_i$ have to satisfy the relations 
\begin{equation}
  \label{comm}
  [P_j,P_4]=[P_j,P_5]=0,\quad j=1,\dots,5,
\end{equation}
and
\begin{equation}
  \label{rel-ll}
  \begin{aligned}
     {}&[P_1,[P_2,P_3]]=[P_2,[P_3,P_1]]=[P_3,[P_1,P_2]]=0,\\ 
     {}&[P_2,[P_2,P_3]]-[P_1,[P_1,P_3]]+(j_1-j_2)P_3=0,\\
     {}&[P_3,[P_3,P_1]]-[P_2,[P_2,P_1]]+(j_2-j_3)P_1=0,\\
     {}&[P_1,[P_1,P_2]]-[P_3,[P_3,P_2]]+(j_3-j_1)P_2=0.
  \end{aligned}
\end{equation}

In the \emph{full anisotropy} case 
\begin{equation}
  \label{j}
  j_1\neq j_2,\ j_2\neq j_3,\ j_3\neq j_1,
\end{equation}
the Lie algebra defined by the generators $P_1,\,P_2,\,P_3$ and
relations \eqref{rel-ll} was described in \cite{ll} explicitly as follows.
Consider the ideal $\mathcal{I}\subset\mathbb{C}[v_1,v_2,v_3]$ generated by the
polynomials 
\begin{equation}
  \label{elc}
  v_\alpha^2-v_\beta^2+j_\alpha-j_\beta,\quad \alpha\,,\beta=1,2,3,
\end{equation}
and set $E=\mathbb{C}[v_1,v_2,v_3]/\mathcal{I}$, i.e., $E$ is the ring of regular
functions on the affine elliptic curve in $\mathbb{C}^3$ defined by polynomials
\eqref{elc}. The image of $v_j\in\mathbb{C}[v_1,v_2,v_3]$ 
in $E$ is denoted by $\bar v_j$.
Consider also a basis $x,\,y,\,z$ of the Lie algebra 
$\mathfrak{sl}_2(\mathbb{C})\cong\mathfrak{so}_3(\mathbb{C})$ 
with the relations 
$$
[x,y]=z,\quad [y,z]=x,\quad [z,x]=y
$$
and endow the space $L=\mathfrak{sl}_2\otimes_\mathbb{C} E$ with 
the natural Lie algebra structure 
\begin{equation*}
  [a\otimes f,\,b\otimes g]=[a,b]\otimes fg,\quad
  a,\,b\in\mathfrak{sl}_2(\mathbb{C}),\ f,\,g\in E. 
\end{equation*}

\begin{proposition}[\cite{ll}]
\label{sll}
Consider the Lie algebra $P$ over $\mathbb{C}$ given by 
generators $P_1,\,P_2,\,P_3$ and relations \eqref{rel-ll}.
Suppose that the numbers $j_1,\,j_2,\,j_3$ are distinct.
Then the mapping 
\begin{equation}
  \label{map}
  P_1\mapsto x\otimes\bar v_1,\quad 
  P_2\mapsto y\otimes\bar v_2,\quad
  P_3\mapsto z\otimes\bar v_3.
\end{equation}
gives an isomorphism of $P$ onto the 
subalgebra $R\subset L$ generated by the elements 
$x\otimes\bar v_1,\,y\otimes\bar v_2,\,z\otimes\bar v_3\in L$. 
\end{proposition}

\section{The prolongation algebra of the nonsingular KN equation}  
\label{pakn}

It is shown in \cite{loop} and rediscovered in \cite{novikov}
that Sklyanin's zero-curvature representation  
$$
D_tM-D_xN+[M,N]=0 
$$ 
for \eqref{l-l} (see \cite{skl,loop,ll} for the precise form of $M,\,N$) 
leads to a zero-curvature representation for the equation
\begin{equation}
  \label{kn1}
  u_t=u_3-\frac32\frac{u_2^2}{u_1}+\frac{bu^4-cu^2+b}{u_1}.
\end{equation}
Namely, denote by $\tilde M=\tilde M(u,u_1)$ 
the matrix function obtained from $M=M(S_1,S_2,S_3)$
by the substitution
\begin{equation}
  \label{sub}
  S_1=\frac{u}{u_1},\quad S_2=i\frac{u^2+1}{2u_1},
  \quad S_3=\frac{u^2-1}{2u_1}.
\end{equation}
Then there is a matrix-function $N'(u,u_1,u_2,u_3)$ such that 
the pair $\tilde M,\,N'$ forms 
a zero-curvature representation for \eqref{kn1}.
Here and below $i=\sqrt{-1}\in\mathbb{C}$.

\begin{remark}
Transformation \eqref{sub} does not map solutions of \eqref{kn1} 
to solutions of the LL equation, since, for instance,  
\begin{equation}
\label{sum=0}
\bigl(\frac{u}{u_1}\bigl)^2+\bigl(i\frac{u^2+1}{2u_1}\bigl)^2+
\bigl(\frac{u^2-1}{2u_1}\bigl)^2=0,
\end{equation}
while in \eqref{l-l} we have $S_1^2+S_2^2+S_3^2=1$.
However, below we use \eqref{sub} to establish one-to-one 
correspondence between prolongation structures 
of the two equations. 

On the other hand, formulas \eqref{sub} and \eqref{sum=0}
inspire one to consider the analog of the LL equation \eqref{l-l}
with the requirement $S_1^2+S_2^2+S_3^2=0$. 
Unfortunately, the prolongation algebra for the
resulting system is trivial, and \eqref{sub} still does not map
solutions of \eqref{kn1} to solutions of this system.   
Perhaps a better understanding of relations between the LL equation
and the KN equation can be derived from the results of \cite{loop}. 
\end{remark}

Motivated by formulas \eqref{All} and \eqref{sub}, we proceed in
describing the prolongation algebra of the KN
equations as follows. Suppose first that \eqref{gkn} takes the form
\begin{equation}
\label{bckn}
u_t=u_3-\frac32\frac{u_2^2}{u_1}+\frac{bu^4-cu^2+b}{u_1}+au_1,\ 
a,\,b,\,c\in\mathbb{C},
\end{equation}
We rewrite \eqref{A1} in the more convenient form
\begin{equation}
\label{A1P}
A_1=uP_1+i\frac{u^2+1}{2}P_2+\frac{u^2-1}{2}P_3,
\end{equation}
i.e., $A_{11}=P_1,\,A_{12}=(iP_2+P_3)/2,\,A_{10}=(iP_2-P_3)/2$.

Evidently, the elements
\begin{equation}
\label{gen_bckn}
	A_0,\ B_0,\ P_1,\ P_2,\ P_3
\end{equation}
represent another set of
generators for the prolongation algebra. Let us write down the
corresponding relations. From \eqref{0} one gets
\begin{equation}
  \label{ab}
  [P_{j},A_0]=[P_{j},B_0]=[A_0,B_0]=0,\ \ j=1,2,3,
\end{equation}
while equation \eqref{-2} gives the following relations between $P_{j}$
\begin{equation}
\label{rel-kn}
\begin{aligned}
{}&[P_1,[P_2,P_3]]=[P_2,[P_3,P_1]]=[P_3,[P_1,P_2]]=0,\\
{}&\frac83 bP_1 = [P_3,[P_3,P_1]]-[P_2,[P_2,P_1]],\\
{}&\frac13(4b+2c)P_2 = -[P_1,[P_1,P_2]]+[P_3,[P_3,P_2]],\\
{}&\frac13(4b-2c)P_3 = [P_1,[P_1,P_3]]-[P_2,[P_2,P_3]]. 
\end{aligned}
\end{equation}
We see that these relations coincide with relations \eqref{rel-ll} for  
\begin{equation}
\label{jbc}
j_2-j_3=-\frac83b,\ \ j_3-j_1=\frac13(4b+2c),\ \ j_1-j_2=\frac13(4b-2c).
\end{equation}
This implies the following theorem.

\begin{theorem}
\label{main}
For any covering of the LL equation \eqref{l-l} 
given by \eqref{All}, \eqref{Bll}  
we obtain a covering   
of equation \eqref{bckn} with \eqref{jbc} as follows.
Substituting \eqref{sub} to \eqref{All}, one gets 
the corresponding $x$-part \eqref{Au_1}, which in turn 
determines the $t$-part by formula \eqref{b},
where $B_0(w)$ is an arbitrary vector field commuting
with the $x$-part, for example one may take $B_0=0$. 

And vice versa, given 
a covering of equation \eqref{bckn} 
with $x$-part \eqref{Au_1}, one obtains
a covering of the form \eqref{All}, \eqref{Bll} 
for the LL equation \eqref{l-l} satisfying \eqref{jbc}.
Namely, the fields $P_1,\,P_2,\,P_3$ 
are determined through decomposition \eqref{A1P},
while $P_4,\,P_5$ are taken such that
they meet \eqref{comm}, for example $P_4=P_5=0$.   
\end{theorem}

An example of this construction
is given in Section \ref{pzcr}.	

It is easily seen that numbers
\eqref{jbc} are nonzero 
if and only if the roots of the polynomial $bu^4-cu^2+b$ are distinct.
In this case the Lie algebra with generators $P_1,\,P_2,\,P_3$ and 
relations \eqref{rel-kn} is described
by Proposition \ref{sll}, which implies the following statement.

\begin{theorem}
\label{main1}	
The prolongation algebra of equation \eqref{bckn} 
with generators \eqref{gen_bckn}
and relations \eqref{ab}, \eqref{rel-kn} 
is isomorphic to the direct sum $C\oplus R$, where
$C=\langle A_0,\,B_0\rangle$ is 
a commutative $2$-dimensional algebra and 
$R$ is the algebra defined in Proposition \ref{sll} 
by means of mapping \eqref{map},
the numbers $j_\alpha-j_\beta$ being given by \eqref{jbc}.
\end{theorem}

Return to the nonsingular KN equation \eqref{kn}. 
Following \cite{novikov}, consider
the linear-fractional transformation 
\begin{equation}
  \label{ft}
  u\mapsto e_1+\frac14(p^4-q^4)
  \frac{q-up}{q+up},
\end{equation}
where $p,\,q\in\mathbb{C}$ are some solutions of the system 
\begin{equation}
\label{pq}
	p^2q^2=e_2-e_3,\quad p^4+q^4=6e_1
\end{equation}
and $e_1,\,e_2,\,e_3\in\mathbb{C}$ are the roots of the polynomial
$4u^3-g_2u-g_3$. 

\begin{lemma}
\label{lem}
Transformation \eqref{ft} is nontrivial if and only if the numbers
$e_1,\,e_2,\,e_3$ 
are distinct, i.e., the KN equation \eqref{kn} is nonsingular. 
In this case transformation 
\eqref{ft} turns equation \eqref{kn} into equation \eqref{bckn} with 
\begin{equation}
\label{bce}
	b=e_2-e_3,\quad c=6e_1.
\end{equation}
\end{lemma}
\begin{proof}
Clearly, transformation \eqref{ft} is nontrivial if and only if 	
\begin{gather}
\label{pq0}
p\neq 0,\quad q\neq 0,\\
\label{p4q4}
p^4-q^4\neq 0.	
\end{gather}
By definition \eqref{pq},  
condition \eqref{pq0} is equivalent to $e_2\neq e_3$.
In addition, we have $(p^4-q^4)^2=4(3e_1+e_2-e_3)(3e_1-e_2+e_3)$,
which implies, taking into account $e_1+e_2+e_3=0$, that 
\eqref{p4q4} is equivalent to $e_1\neq e_2,\,e_1\neq e_3$.
This proves the first statement of the lemma, while the second
statement is straightforward to check.  
\end{proof}

\begin{remark}
Existence of a linear-fractional transformations from \eqref{kn} 
to \eqref{bckn} 
was claimed already in \cite{loop}, but a formula was not given there.     
Note that the class of equations \eqref{gkn} and 
the form \eqref{Au_1}, \eqref{A1} 
of the prolongation structure are preserved by arbitrary 
linear-fractional transformations $u\mapsto (k_1u+k_2)/(k_3u+k_4)$.
\end{remark}

Since isomorphic equations have the same prolongation algebra,
Lemma \ref{lem} implies that the prolongation algebra of 
the nonsingular KN equation \eqref{kn} 
is isomorphic to the prolongation algebra
of equation \eqref{bckn} with \eqref{bce}
described in Theorem \ref{main1} and Proposition \ref{sll}.
In this case polynomials \eqref{elc} are 
\begin{equation}
\label{elc_kn}
v_i^2-v_j^2+\frac83(e_j-e_i),\quad i,\,j=1,2,3.
\end{equation}
\begin{theorem}
\label{main2}
Let $\mathcal{I}\subset\mathbb{C}[v_1,v_2,v_3]$ be the ideal generated
by polynomials \eqref{elc_kn} and 
$E=\mathbb{C}[v_1,v_2,v_3]/\mathcal{I}$.
The prolongation algebra of 
the nonsingular KN equation \eqref{kn} 
is isomorphic to the 
direct sum $C\oplus R$, where
$C$ is a commutative $2$-dimensional algebra and 
$R$ is the subalgebra of $\mathfrak{sl}_2\otimes_\mathbb{C} E$ 
generated by 
$x\otimes\bar v_1,\,y\otimes\bar v_2,\,z\otimes\bar v_3$.
\end{theorem}

The explicit isomorphism is derived through transformation
\eqref{ft} from the one for equation \eqref{bckn} described
in Theorem \ref{main1}. 

\begin{remark} 
The known for \eqref{kn} $\mathfrak{sl}_2$-valued ZCR  
with elliptic parameters \cite{loop,novikov} 
arises from the restriction to $R$ of the family of homomorphisms 
$$
\rho_a\colon\mathfrak{sl}_2\otimes_\mathbb{C} E\to\mathfrak{sl}_2,\quad 
x\otimes p\mapsto p(a)x,
$$ 
parameterized by the points $a\in\mathbb{C}^3$ of the elliptic curve. 
\end{remark}

\section{A polynomial zero-curvature representation}
\label{pzcr}

In particular, Theorem \ref{main} 
establishes one-to-one correspondence
between ZCRs for the LL equation \eqref{l-l} 
and equation \eqref{bckn} with \eqref{jbc}. 
Consider the following example. 

In \cite{polyn}  
a zero-curvature representation 
$D_xM-D_tN+[M,N]=0$ was found for the LL equation \eqref{l-l} with 
\begin{equation}
\label{M}
	M=\frac12 S(\lambda+\tilde J)
\end{equation}
(the form of $N$ is not important for us), where
\begin{gather*}
S=\frac12
	\begin{pmatrix}
	0 & S_1 & S_2 & S_3\\
	-S_1 & 0 & S_3 & -S_2\\
	-S_2 & -S_3 & 0 & S_1\\
	-S_3 & S_2 & -S_1 & 0
	\end{pmatrix},\\
	\tilde J=\mathrm{diag}
	(-j_1'-j_2'+j_3',\,-j_1'+j_2'-j_3',\,j_1'-j_2'-j_3',\,
	j_1'+j_2'+j_3'),\\
	j_k'=\sqrt{-4j_k},\quad k=1,\,2,\,3,
\end{gather*}
and $\lambda$ is an unconstrained complex parameter.

Denote by $\tilde M\in\mathfrak{sl}_4(\mathbb{C})$ the matrix obtained
from \eqref{M} by substitution \eqref{sub}.
By formulas \eqref{b}, \eqref{ABM} and the procedure of 
Theorem \ref{main}, the matrices $\tilde M$ and  
\begin{multline*}
  N'=-\frac{u_3}{u_1}\tilde M+ 
  \frac{u_2^2}{2u_1^2}\tilde M+
  2u_2\frac{\partial \tilde M}{\partial u}\\
  -\frac{bu^4-cu^2+b}{3u_1^2}\tilde M-2u_1[\tilde M,\frac{\partial \tilde M}{\partial u}]
  -2u_1^2\frac{\partial^2 \tilde M}{\partial u^2}+a\tilde M,
\end{multline*}
constitute a ZCR $D_t\tilde M-D_x N'+[\tilde M,N']=0$
for \eqref{bckn} with \eqref{jbc}.

It is easily seen that for any $b,\,c\in\mathbb{C}$ there exist
$j_1,\,j_2,\,j_3\in\mathbb{C}$ such that \eqref{jbc} holds.
Therefore, we have constructed a ZCR for each equation \eqref{bckn}. 
Applying transformation \eqref{ft}, we obtain a new ZCR 
for the nonsingular KN equation \eqref{kn}.
 
Interestingly, 
this ZCR is polynomial in the spectral parameter 
in contrast to the known for \eqref{kn} ZCRs with elliptic 
parameters \cite{loop,krich,novikov}.

\section*{Acknowledgments}
The authors thank I.~S.~Krasilshchik and P.~H.~M.~Kersten for
useful discussions.


\begin{thebibliography}{99}

\bibitem{adler}
V.~E.~Adler. 
B\"acklund transformation for the Krichever-Novikov equation. 
\emph{Internat. Math. Res. Notices} \textbf{1998} no.~1, 1--4.

\bibitem{rb}
A.~V.~Bocharov, V.~N.~Chetverikov, S.~V.~Duzhin, N.~G.~Khor{\cprime}kova,
I.~S.~Krasil{\cprime}shchik, A.~V.~Samokhin, Yu.~N.~Torkhov, 
A.~M.~Verbovetsky, and  A.~M.~Vinogradov.
\emph{Symmetries and Conservation Laws for Differential Equations of
Mathematical Physics}.
Amer. Math. Soc., Providence, RI, 1999.


\bibitem{polyn}
L.~A.~Bordag and A.~B.~Yanovski.
Polynomial Lax pairs for the chiral $O(3)$-field equations and the 
Landau-Lifshitz equation.
\emph{J. Phys. A: Math. Gen.} \textbf{28} (1995), 4007--4013. 


\bibitem{loop}
F.~Guil and M.~Ma\~nas. 
Loop algebras and the Krichever-Novikov equation. 
\emph{Phys. Lett. A} \textbf{153} (1991), 90--94. 


\bibitem{nonl} 
I.~S.~Krasilshchik and A.~M.~Vinogradov. 
Nonlocal trends in the geometry of differential equations. 
\emph{Acta Appl. Math.} \textbf{15} (1989), 161--209.

\bibitem{krich} 
I.~M.~Krichever and  S.~P.~Novikov. 
Holomorphic bundles over algebraic curves, and nonlinear equations. 
\emph{Russian Math. Surveys} \textbf{35} (1980), 53--80. 

\bibitem{novikov}
D.~P.~Novikov. 
Algebro-geometric solutions of the Krichever-Novikov equation. 
\emph{Theoret. and Math. Phys.} \textbf{121} (1999), 1567--1573. 

\bibitem{pkdv}
G.~H.~M.~Roelofs and R.~Martini.
Prolongation structure of the KdV equation in the bilinear form of
Hirota. \emph{J. Phys. A: Math. Gen.} \textbf{23} (1990), 1877--1884.

\bibitem{ll}  
G.~H.~M.~Roelofs and R.~Martini.
Prolongation structure of the Landau-Lifshitz equation.
\emph{J. Math. Phys.} \textbf{34} (1993), 2394--2399. 


\bibitem{skl}
E.~K.~Sklyanin. On complete integrability of the Landau-Lifshitz equation.
\emph{Preprint LOMI E-3-79}, Leningrad, 1979.

\bibitem{sok1}
S.~I.~Svinolupov and V.~V.~Sokolov.
Evolution equations with nontrivial conservation laws. 
\emph{Functional Anal. Appl.} \textbf{16} (1982), 317--319.

\bibitem{sok2}
S.~I.~Svinolupov, V.~V.~Sokolov, and R.~I.~Yamilov. 
B\"acklund transformations for integrable evolution equations.
\emph{Soviet Math. Dokl.} \textbf{28} (1983), 165--168.

\bibitem{kdv_pa} 
H.~N.~van Eck. 
The explicit form of the Lie algebra of Wahlquist and
Estabrook. A presentation problem. 
\emph{Nederl. Akad. Wetensch. Indag. Math.} \textbf{45}
(1983), 149--164.      

\bibitem{prol} 
H.~D.~Wahlquist and F.~B.~Estabrook. 
Prolongation structures of nonlinear evolution equations. 
\emph{J. Math. Phys.} \textbf{16} (1975), 1--7.

\end{thebibliography}
\end{document}